\documentclass[twocolumn,prb,aps]{revtex4}
\usepackage{graphicx}% Include figure files
\usepackage{dcolumn}% Align table columns on decimal point
\usepackage{bm}% bold math
\begin{document}
\title{Angular dependence of domain wall resistivity in SrRuO$_{{\bf 3}}$ films.}
\author{Michael Feigenson}
\author{Lior Klein}
\affiliation{Physics Department, Bar Ilan University, Ramat Gan 52900, Israel}
\author{James W. Reiner} 
\altaffiliation{Current location: Department of Applied Physics,
Yale University, New haven, Connecticut 06520-8284}
\author{Malcolm R. Beasley}
\affiliation{T. H. Geballe Laboratory for Advanced Materials, Stanford University,
Stanford, California 94305}

\date{\today}


\begin{abstract}
${\rm SrRuO_3}$ is a 4d itinerant ferromagnet (T$_{c}$ $\sim $150 K) with stripe
domain structure. Using high-quality thin films of SrRuO$_{3}$ we study the
resistivity induced by its very narrow ($\sim 3$ nm) Bloch domain walls, $\rho_{DW}$ (DWR), at
temperatures between 2 K and T$_{c}$ as a function of the angle, $\theta $, between the
electric current and the ferromagnetic domains walls. We find that  
$\rho_{DW}(T,\theta)=\sin^2\theta \rho_{DW}(T,90)+B(\theta)\rho_{DW}(T,0)$ which 
provides the first experimental indication that the angular dependence of spin accumulation
contribution to DWR is $\sin^2\theta$. We expect magnetic multilayers to exhibit a similar behavior.

\end{abstract}

\maketitle

\section{Introduction}
Following the discovery of the giant magnetoresistance (GMR) effect in
magnetic multilayers \cite{gmr} and related phenomena considered to lay the
foundation for the emerging field of spintronics,\cite{Spintronics} there
has been an intensive effort to elucidate the mechanisms involved in spin
polarized transport in the presence of magnetic interfaces. One of the
outcomes of this effort was the realization that naturally obtained domain
structure in itinerant ferromagnets may offer an important opportunity to study basic
issues relevant to spintronics while avoiding difficulties encountered 
when artificially grown multilayers are
studied. These difficulties arise due to the intrinsic uncertainties
and complications associated with the nature of magnetic interfaces in these systems. 

Two major hurdles were encountered along this
route when domain wall resistivity (DWR) of iron,\cite{Kent} nickel,\cite{Viret} cobalt\cite{Gregg,Viret} 
and FePd\cite{Ravelosona}
was studied: (a) The resistivity (or interface resistance)  is quite small and hardly distinguished
from the anisotropic magnetoresistance effect which is unavoidable in these systems due to the presence
of closure domains.\cite{Rudiger prb}      (b) The estimated width of ferromagnetic
domain walls (DWs) in these systems 
is relatively large (e.g., 10 nm in FePd,\cite{Ravelosona} 
15 in cobalt and 100 in nickel\cite{Viret}). Consequently,
models applicable to magnetic multilayers  with 
atomically sharp interfaces are not relevant and
other models were used to interpret the obtained results.\cite{Tatara,Levy,Gorkom} Therefore, limited
 understanding of the physics governing magnetic multilayers could be achieved from studying DWR
in such systems.

These problems are absent in the itinerant ferromagnet  SrRuO$_{3}$. 
This compound has relatively large
magnetocrystalline anisotropy field ($\sim 10 \ {\rm T}$) 
that yields stripe structure with domain wall separation of 200 nm 
  without closure domains and with narrow
Bloch DWs whose estimated width is only $\delta
\sim  3 \ {\rm nm}$. These features make SrRuO$_{3}$ a model system
on which models suggested for magnetic multilayers can be tested.

Initial evidence that the same physical mechanisms are relevant both to
DWR in  SrRuO$_{3}$ and magnetic multilayers  was given in a previous report \cite{Klein prl} where
the magnitude of the DWR with perpendicular current was shown to be qualitatively
and quantitatively consistent
with resistivity induced by two sources relevant to magnetic multilayers: spin accumulation 
and potential step. We should note that spin accumulation was also claimed to be
the source of DWR in
nanowires of cobalt\cite{ebels}; however, this attribution was later challenged on grounds that the effect
is expected to be suppressed in the 15 nm thick DWs.\cite{simanek}

Here we explore DWR in SrRuO$_{3}$ 
as a function of the angle, $\theta $, between
the electric current and the ferromagnetic DWs. We find that 
DWR for any angle $\theta$  is given by a $temperature-independent$ linear combination
of DWR for parallel ($\rho_{DW}(T,\theta=0)$) 
and perpendicular ($\rho_{DW}(T,\theta=90)$) currents.
As we demonstrate below, this fit provides the first
experimental indication that the angular dependence of spin accumulation
contribution  to DWR is proportional to $\sin^2\theta$, as expected based on simple theoretical
considerations. This behavior  is   likely to be found in magnetic
multilayers, as well.

\section{Experiment}
SrRuO$_{3}$ is a pseudocubic perovskite and an itinerant ferromagnet
with Curie temperature of ${\rm \sim160 \ K}$ for bulk and ${\rm \sim150 \ K}$ for thin films (provided
the film thickness is more
than 10 nm). Thin films exhibit high magnetocrystalline anisotropy field (${\rm \sim 10 \ T}$)
with uniaxial anisotropy with the easy axis roughly\cite{easy axis} along the crystallographic $b$
axis in the orthorhombic notation.  For this study we used high-quality thin films of SrRuO$_{3}$
grown on 
${\rm SrTiO_3}$ substrates by reactive electron beam coevaporation.\cite{Benerofe}
 To avoid twinning in the SrRuO$_{3}$ film and obtain
a film with the same orientation of
the uniaxial anisotropy throughout the sample, the cubic symmetry of the substrate
surface needs to be broken. This is achieved by slightly miscutting
($\sim  2$ degrees) the  ${\rm SrTiO_3}$ substrates which forms atomically flat terraces separated
by unit cell steps. The
film whose growth starts  at the steps grows uniformly with the projection of the easy axes on the plane of
the film perpendicular to the miscut-induced steps. 

The domain structure of similarly grown films was thoroughly studied using
Lorentz microscopy on free standing films after removing the ${\rm SrTiO_3}$
substrate with a chemical etch.\cite{Marshall} It was found that in the
domain state
magnetic stripes are formed  parallel to the easy axis projection on the film.
Figure 1 shows DWs image taken from Ref.\onlinecite{Klein prl}. The DWs are 200 nm apart
 with no detectable dependence on film thickness (in the studied range of 30 nm - 100 nm) and 
no detectable
dependence on temperature except for few degrees below ${\rm T_c}$.
The estimated thickness of the Bloch wall,  $\delta$,  is  $\sim 3$ nm.\cite{wall} 
Closure domains were not observed as expected for SrRuO$_{3}$ whose
  $Q = \frac{K}{2\pi M_{s}} \gg 1$
(here K is the anisotropy energy and $M_s$ is the saturated magnetization).
The stripe structure forms spontaneously when the sample is cooled below ${\rm T_c}$ in zero field. When a
sufficiently high field is applied below ${\rm T_c}$ the magnetization becomes uniform with no stripes. An
important observation is that the uniform magnetization state remains stable when the field is set back to
zero. A substantial (temperature dependent) field needs to be applied in the negative direction to start
magnetization reversal (with many fewer DWs than in the initial zero-field-cooled state). These features are
essential for facilitating clear identification of DWR in this system.

\begin{figure}
\includegraphics[scale=.6]{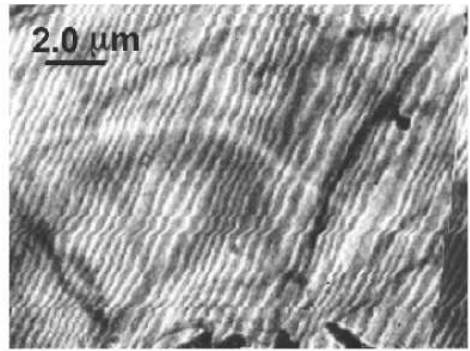}
\caption{(a) Image
of stripe domain walls in $\rm SrRuO_3$ with transmission electron microscopy
 in Lorentz mode (from Ref. \onlinecite{Klein prl}). The bright and dark lines image walls that diverge or
converge the electron beam, respectively. Background features are related to buckling of the free standing
film and are not related to magnetic variations. 
The average spacing between the walls is $\sim 200 \ \rm nm$.
(b) Schematic figure of the photolitography patterns used 
for angular dependence study. The patterns are at eight different orientations relative to the DWs (marked
by the arrow):
0, 15, 30, 45, -45, 60, 75 and 90 degrees.}
\end{figure}

Using photolithography we patterned films in the form shown in Figure 1.
It includes patterns at eight different angles relative to the stripes: 0,
15, 30, 45, -45, 60, 75 and 90 degrees. The distance between the
two distant voltage leads is 500 $\mu m$ and the width of the current path in the
patterns is 50 $\mu m$. Therefore, for $\theta $\ = 90 the current crosses $%
\sim $ 2500 domain walls between the voltage leads, and for $\theta $ =
0 there are $\sim $ 250 parallel DWs running along the current path.

\begin{figure}
\label{dwrvt}
\includegraphics[scale=0.8]{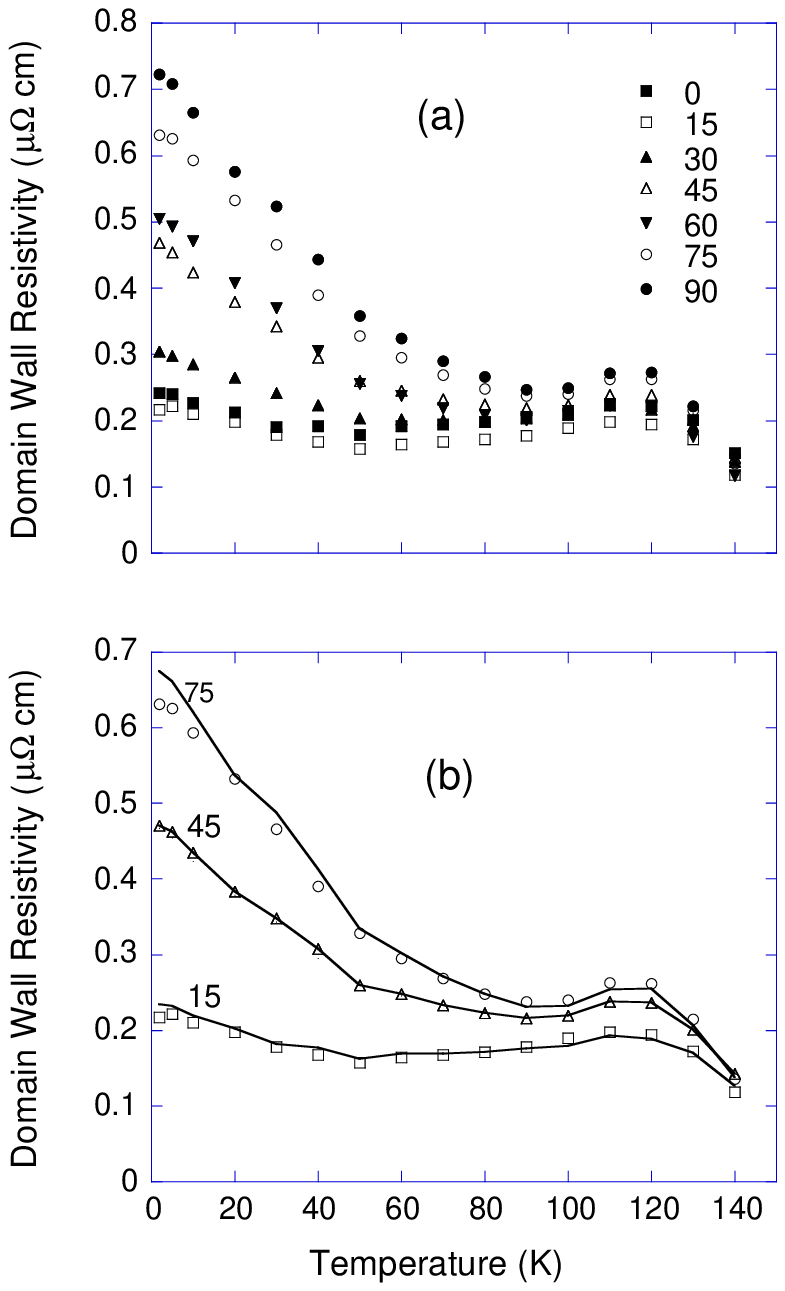}
\caption{ (a): DWR as a function of temperature with current flowing at different
angles relative to the DWs: $\theta$=0, 15, 30, 45, 60, 75 and 90 degrees.
The data points for $\theta$=-45 are on top of the data points for $\theta$=45 and are
not shown in the figure (b) DWR as a function
of temperature with  $\theta$=15,  45, and 75. The symbols are the actual data points
for the different angles whereas the line is the one-parameter fit based on Equation 1.}
\end{figure}

The DWR measurements were performed for each film on all eight patterns at
temperatures between 2 K and 140 K. To measure the excess resistivity
induced by domain walls at a specific temperature T$_{measure}$, we cooled
the sample in zero magnetic field from above T$_{c}$ down to T$_{measure}$
and measured the resistivity there. As noted before,\cite{Klein prl}
when cooling in zero field a stripe domain structure forms; therefore, the zero-field-cooled
resistivity is measured with the magnetic domains present. While staying at
the same temperature we increased the field to obtain uniform magnetization in
the film. We then decreased the field to zero and measured the resistivity
once again. The uniform magnetization remains stable at zero field. Therefore, the second
 resistivity is  measured in the absence of any DWs. Consequently, we can
attribute the difference between the two values of resistivity to DWR.
For each temperature where  DWR was measured, the process was repeated by
first warming the sample above ${\rm T_c}$ and cooling in zero field to the
temperature where DWR was to be measured.

\section{Results and discussion}
Fig. 2a shows the temperature dependence of DWR for the current
flowing in eight different angles relative to the DWs. Our main objective is to understand the
measured DWR changes as a function of the angle between the current and the DWs.

As noted
before, we analyze our results by applying models used for magnetic
multilayers for which two important contributions to resistivity  were considered theoretically: spin
accumulation\cite{Valet} and potential step.\cite{Barnas} 
For parameters appropriate for ${\rm SrRuO_3}$ the two effects are expected
to yield interface resistance for perpendicular current on the order of 
10$^{-15}$ ${\rm \Omega \ m^{2}}$, as experimentally observed.\cite{Klein prl}

Spin accumulation is generated when spin-polarized current crosses an interface
between two domains with opposite magnetization. Valet and Fert\cite{Valet} showed that such a
current yields spin accumulation near the interface that induces a potential
barrier which results in interface resistance $r$ given by $r=2\beta^2 \rho_Fl_{sf}$ (see Eq. 25
in Ref. \onlinecite{Valet}) where $\beta$ is the spin asymmetry coefficient (zero
for unpolarized current and 1 for fully polarized current), $\rho_F$ is the average resistivity
(for spin-up and spin-down currents) and  $l_{sf}$ is the spin diffusion length which is the
characteristic length over which the polarization of crossing current equilibrates with the
equilibrium polarization. The spin-accumulation resistivity, $\rho_{DW}^{SA}$, is proportional
to the interface resistance, r, with the number  of domain walls
per unit length being the proportionality factor. $\rho_{DW}^{SA}$ is expected to
decrease with temperature due to the expected fast decrease in $l_{sf}$ due to increase in magnetic
scattering (e.g., due to magnons) which becomes more probable with increasing temperature. 

Since spin accumulation is associated with the net current
crossing the interface, no spin accumulation contribution to DWR is expected
for current parallel to the domain walls. Here we explore in what way spin accumulation
contribution, that we note by
$\rho_{DW}^{SA}$,
changes as a function of angle. 

Two factors relevant to spin accumulation resistivity change as a function
of angle (see  inset to Figure 3a): the net current flowing perpendicular to the interface
is multiplied by $\sin \theta$ and the number of domain walls per unit length crossed
by a current is multiplied by another factor of $\sin \theta$. 
The bottom line is that spin accumulation contribution to DWR for
any angle $\theta$ is expected to be the spin accumulation resistivity for
perpendicular current multiplied by $\sin^2\theta$. 
The problem, however, is to determine the spin
accumulation part in the DWR with perpendicular current.

As we noted before, no spin accumulation 
contribution is expected for parallel current. Therefore, assuming 
that the sources responsible for DWR for parallel current (presumably, related to potential steps) are
present for other angles as well  and assuming that (similarly to spin-accumulation) the contribution
at other angles is  a $temperature-independent$ function of the angle alone, $A(\theta$),  we can
expect that DWR for any temperature and angle will be given by the following equation:

\begin{eqnarray}
\label{the equation}
%\begin{tabular} {lll}
\rho_{DW}(T,\theta)=&&\sin^2\theta
(\rho_{DW}(T,90)-A(90)\rho_{DW}(T,0)) \nonumber\\
 +&&
A(\theta)\rho_{DW}(T,0) \nonumber\\
=&&\sin^2\theta \rho_{DW}(T,90)+ B(\theta) \rho_{DW}(T,0)
%\end{tabular} 
\end{eqnarray}

The term $(\rho_{DW}(T,90)-A(90) \rho_{DW}(T,0))$ is the spin accumulation part in the DWR
for perpendicular current and $B(\theta)=A(\theta)-A(90)\sin^2\theta$. 

To test this model we check whether Equation \ref{the equation} can
reproduce, using  measured $\rho_{DW}(T,90)$ and $\rho_{DW}(T,0)$, the measured DWR at all other
angles  by using a $single$ fitting function, $B(\theta )$. The success of
this fit is visible in Figure 2b.

\begin{figure}
\includegraphics[scale=0.7]{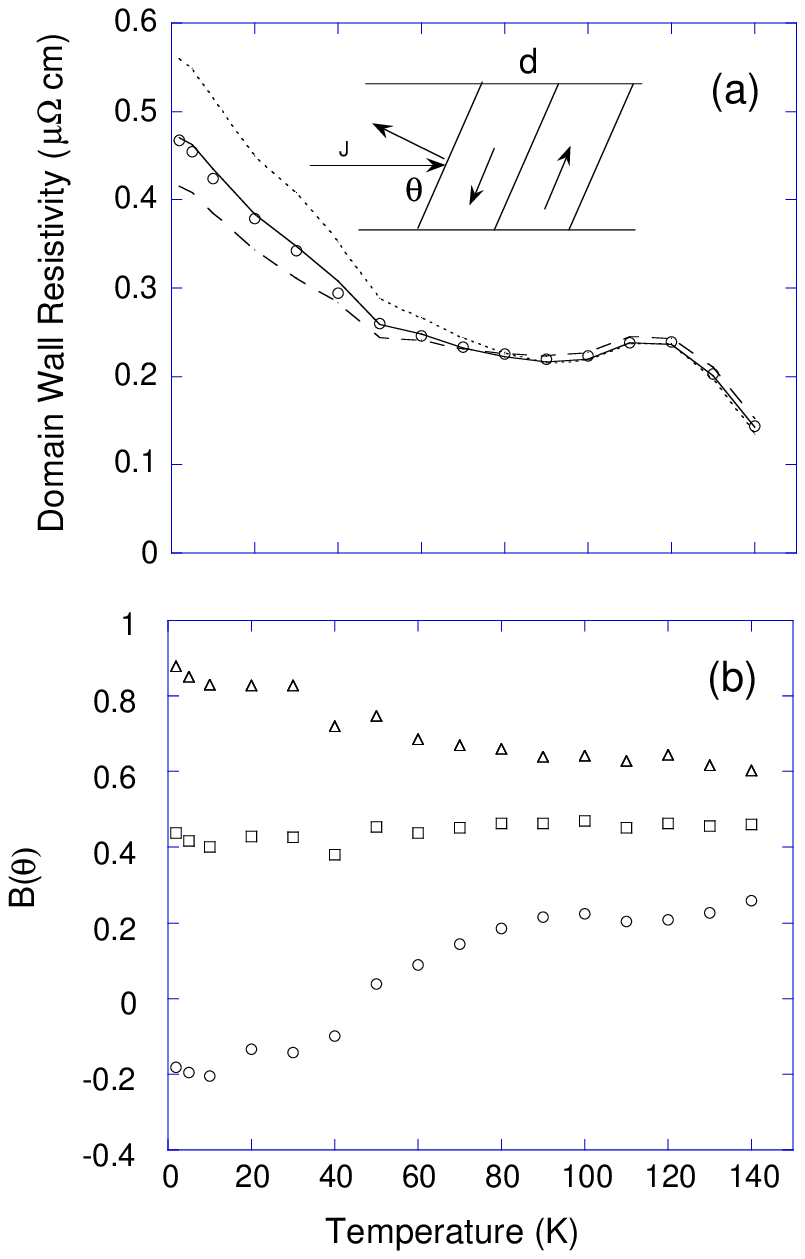}
\caption{(a)  $\rho_{DW}(T,45)$ with three different fits assuming different angular
dependence of the spin accumulation contribution to DWR: $\sin\theta$ (dotted), $\sin^2\theta$ (full)
and $\sin^3\theta$ (dashed). Inset: Schematic illustration of a current path with current $J$ at an angle
$\theta$ relative to the DWs. The current flowing perpendicular to the wall is $J\sin\theta$
and the distance the current flows between the walls $d$ is (domain width)/$\sin\theta$.
(b) $(\rho_{DW}(T,45)-f(\theta)\rho_{DW}(T,90))/\rho_{DW}(T,00)$
as a function of $\theta$ with $f(\theta )=\sin\theta$ (circles), $f(\theta )=\sin^2\theta$ (squares)
and $f(\theta)=\sin^3\theta$ (triangles). Following Equation 1, the quality of the fit is manifested
in temperature independence of $(\rho_{DW}(T,45)-f(\theta)\rho_{DW}(T,90))/\rho_{DW}(T,00)$
}
\end{figure}

As a sensitivity check on our assumption
 of the $\sin^2\theta$ dependence of $\rho_{DW}^{SA}$,  we looked for the best fit (allowing
$B(\theta )$ to vary) when
$\sin^2\theta$ is  replaced by  $\sin\theta$  or $\sin^3\theta$.
 The different fits are shown in Figure 3a for
$\theta = 45$. Figure 3b shows an alternative comparison of the three
fits which is independent of any fitting parameter. It shows
$(\rho_{DW}(T,45)-f(\theta)\rho_{DW}(T,90))/\rho_{DW}(T,0)$ as a function of temperature
with $f(\theta)$ being $\sin\theta$,  
 $\sin^2\theta$, or $\sin^3\theta$. A fit consistent with the assumption that the
sources of DWR for parallel current (that do not include spin accumulation) are
present at other angles as well and their relative effect at other angles is temperature
independent should yield a temperature independent curve. Figure 3b 
demonstrates that the best fit is indeed obtained with $f(\theta)=\sin^2\theta$.

\begin{figure}
\includegraphics[scale=0.8]{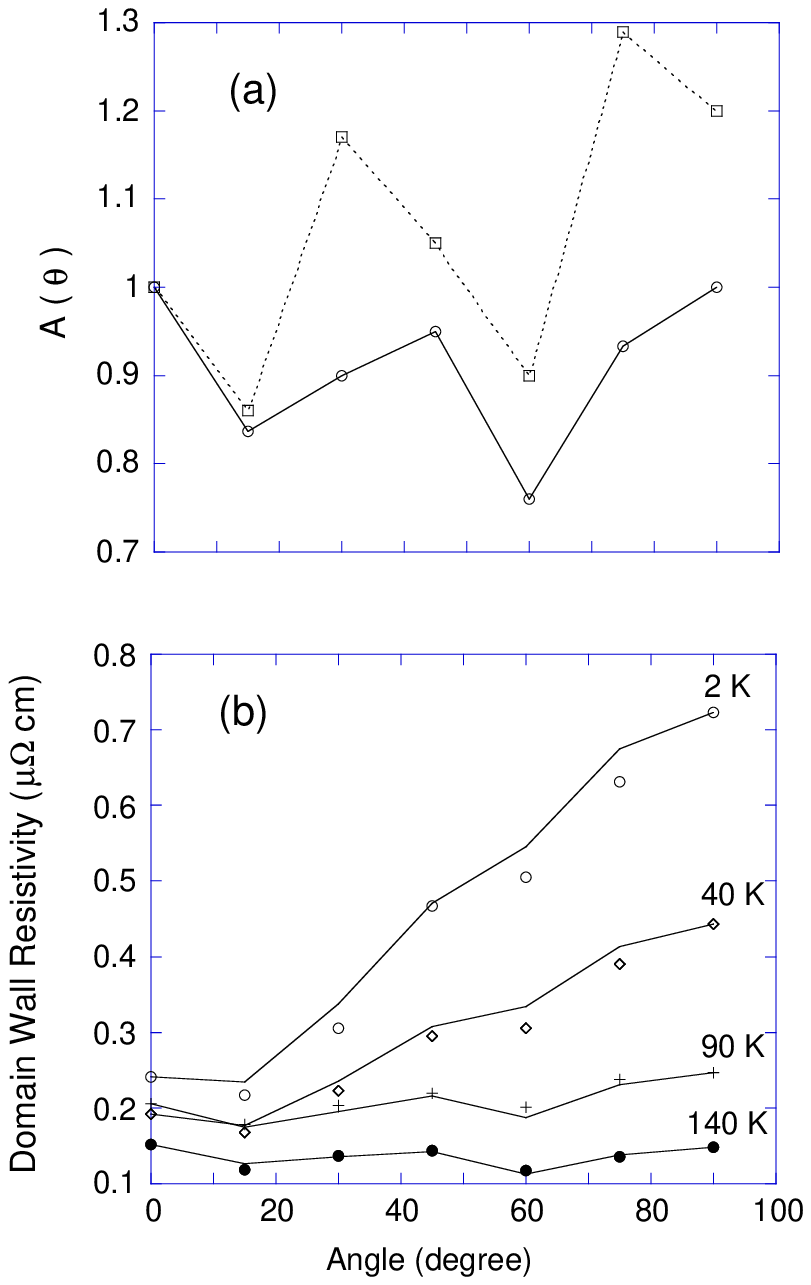}
\caption{(a) $A(\theta)$ as a function of angle for two different samples
(b) $\rho_{DW}(T,\theta)$ as a function of angle at
different temperatures. The symbols are the actual data
points whereas the line is the one parameter fit based on
Equation 1.}
\end{figure}

The fitting function is $B(\theta)=A(\theta)-A(90)\sin^2\theta$. To identify the
spin accumulation part  in the DWR of the perpendicular current we need to determine $A(90)$. 
To identify $A(90)$ we consider that the closer we are to ${\rm T_c}$ the less is 
the relative contribution of spin accumulation to DWR; hence, the angular dependence of DWR is
dominated by the angular dependence of $A(\theta)$. Therefore, we look for  $A(90)$ for
which
$A(\theta)$ is the most similar to the angular dependence of DWR obtained at high temperature.
Figure 4a presents $A(\theta)$ for two different samples and Figure 4b shows the angular
dependence of the DWR at different temperatures.
The main feature 
is the non monotonic or even oscillatory (note correlation between the two samples) 
behavior of $A(\theta)$
which is reflected in the angular dependence of the DWR particularly
near ${\rm T_c}$ where $\rho_{DW}^{SA}$ diminishes. It is likely that the source
that displays this behavior is related to interface resistance associated with potential steps; however,
the specific cause of this behavior is unclear at the moment. It is important
to note that the exact form of $A(\theta)$ depends on the thickness of the film so we
hope that more study focused on the thickness dependence of DWR will yield more
understanding. 

\begin{figure}
\includegraphics[scale=0.5]{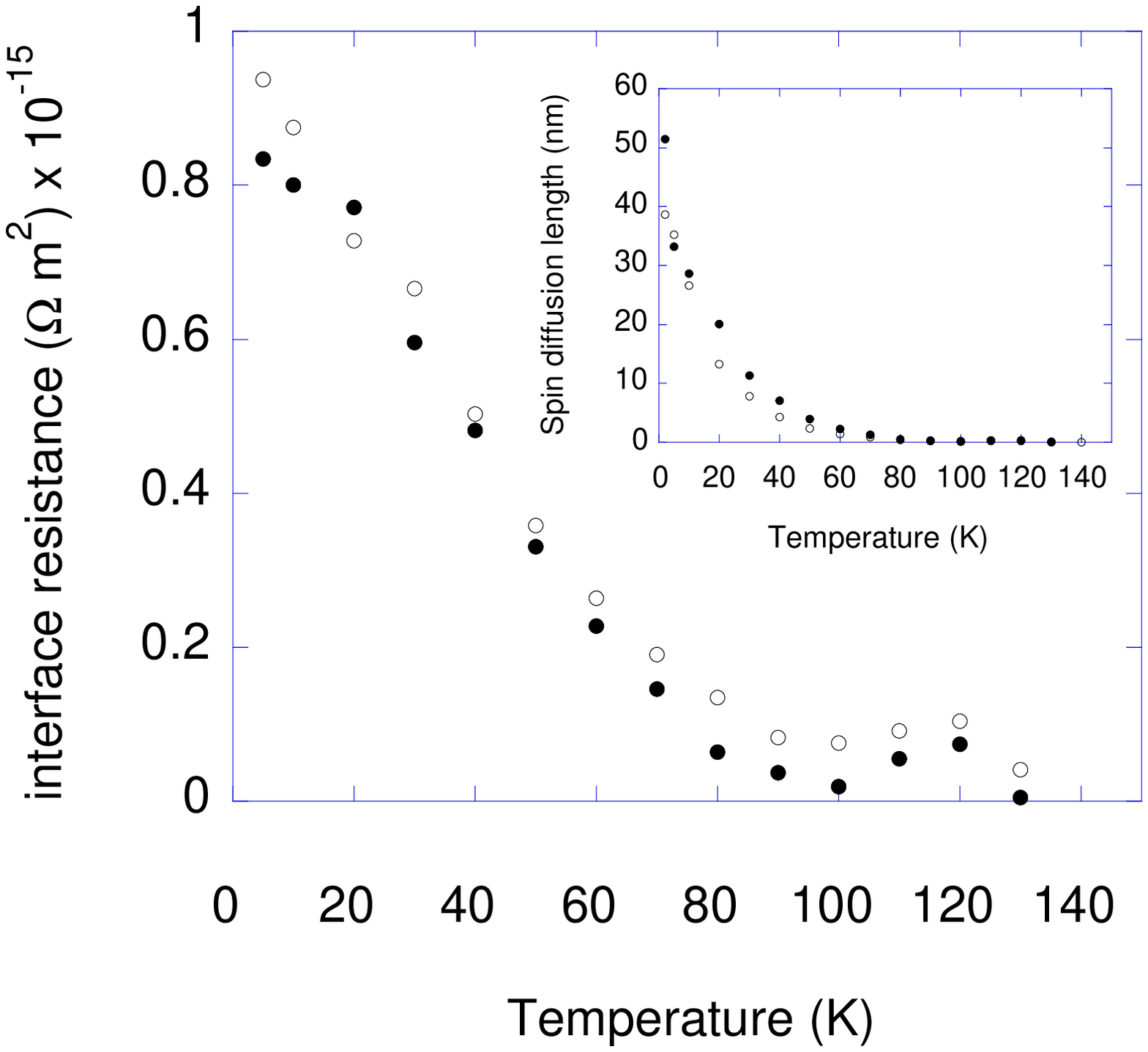} 
\caption{Spin accumulation contribution to the interface resistance as a function of temperature for two
different samples. The inset shows the extracted spin diffusion length based on Valet-Fert\cite{Valet}
relation.}
\end{figure}

Having found $A(90)$, we can determine the spin accumulation interface resistance as a
function of temperature. Figure 5 shows the spin accumulation interface resistance for two different
samples and the inset shows their extracted $l_{sf}$ using Valet-Fert equation.\cite{Valet} 
It is important to note that this derivation of
$l_{sf}$ is expected to be reliable only in the low-temperature limit where Valet-Fert equation is valid. 
From the low temperature limit we find $l_{sf}$ on the order of 40-50 nm. This value is consistent with our
findings that DWR for perpendicular current scales with the density of DWs\cite{dws} which implies
that the scattering at neighboring DWs is independent, suggesting that spin diffusion length is smaller
than the separation between the DWs (200 nm).

\section{Summary and conclusion}
 We present here for the first time data on the angular dependence of
resistivity induced by stripe domain structure in a system that can adequately serve as a model system
for magnetic multilayers. We find a simple fitting equation whose
success provides the first  experimental indication that the angular dependence of 
the spin-accumulation resistivity $\rho_{DW}^{SA}$ is $\sin^2\theta$.

\begin{acknowledgments}
L.K. acknowledges support by the Israel Science Foundation founded by the Israel Academy of
Sciences and Humanities.
\end{acknowledgments}


\begin{thebibliography}{9999}

\bibitem{gmr}  M. N. Baibich {\it et al.}, Phys. Rev. Lett. {\bf 61}, 2472 (1988).

\bibitem{Spintronics} S. A. Wolf {\it et al.}, Science {\bf 294}, 1488 (2001).

\bibitem{Kent}  A. D. Kent {\it et al.}, J. Appl. Phys. {\bf 85}, 5243 (1999).

\bibitem{Viret}  M. Viret {\it et al.}, Phys. Rev. B {\bf 53}, 8464 (1996).

\bibitem{Gregg}  J. F. Gregg {\it et al.}, Phys. Rev. Lett. {\bf 77}, 1580 (1996).

\bibitem{Ravelosona}  D. Ravelosona {\it et al.}, Phys. Rev. B {\bf 59}, 4322 (1999).

\bibitem{Rudiger prb}  U. Rudiger {\it et al.}, Phys. Rev. B {\bf 59} 11914 (1999).

\bibitem{Tatara}  G. Tatara and H. Fukuyama, Phys. Rev. Lett. {\bf 79}, 5110 (1997).

\bibitem{Levy}  P. M. Levy and Zhang, Phys. Rev. Lett. {\bf 78}, 3773 (1997).

\bibitem{Gorkom}  R. P. van Gorkom, A. Brataas and G.E. W. Bauer, Phys.
Rev. Lett. {\bf 83}, 4401 (1999).

\bibitem{Klein prl}  L. Klein {\it et al.}, Phys. Rev. Lett. {\bf 84}, 6090 (2000).

\bibitem{ebels} U. Ebels {\it et al.}, Phys. Rev. Lett. {\bf 84}, 983 (2000).

\bibitem{simanek} E. Simanek Phys. Rev. B {\bf 63}, 224412 (2001).

\bibitem{easy axis} The easy axis is along the $b$ direction near ${\rm T_c}$; however,
there is a reorientation transition where the easy axis changes is orientation from 45 degrees
to the film normal at ${\rm T_c}$ to 30 degrees to the normal in the zero-temperature limit.

\bibitem{Benerofe}  S. J. Benerofe {\it et al.}, J. Vac. Sci Tecjnol. B {\bf 12}, 1217 (1994).

\bibitem{Marshall}  A.F. Marshall {\it et al.}, J.Appl. Phys {\bf 85}, 4131 (1999).

\bibitem{wall} Based on the known relation 
$\delta $ = $\pi \left( \frac{C}{2K_{1}}\right) ^{1/2}$ where K$_{1}$ is the anisotropy constant and C =
$\frac{2JS_{2}}{a}$. Here $J$ is exchange energy, $S$ is spin and $a$ is distance between spins.

\bibitem{Valet}  T. Valet and A.Fert, Phys. Rev. B {\bf 48}, 7099 (1993).

\bibitem{Barnas}  J. Barnas and A.Fert, Phys. Rev. B {\bf 49}, 12835 (1994).

\bibitem{dws} L. Klein {\it et al.}, J. Magn. Magn. Mater. {\bf 226-230}, 780-781 (2001).

\end{thebibliography}
\end{document}